\documentclass[]{hdsr}

\begin{document}

\newgeometry{bottom=1.5in}


\begin{center}

  \title{Quantum Computation, Data Science, and Bell games}

  \maketitle

  \thispagestyle{empty}
  
  \vspace*{.2in}

  \begin{tabular}{cc}
    Richard D. Gill\upstairs{\affilone,*}
   \\[0.25ex]
   {\small \upstairs{\affilone} Mathematical Institute, Leiden University} \\
  \end{tabular}
  
  \emails{
    \upstairs{*}gill@math.leidenuniv.nl 
    }
  \vspace*{0.4in}

\begin{abstract}
I draw attention to statistical and probabilistic and computer science aspects of the very, very close by topic of quantum internet.
\end{abstract}
\end{center}

\vspace*{0.15in}
\hspace{10pt}
  
\copyrightnotice

\section*{Media Summary}
Quantum computing seems to be on the way, but still there are many hurdles to overcome and still there are some who have fundamental doubts that it will ever deliver the goods. In the meantime, a parallel research effort is being made in designing and implementing quantum internet. How about the prospect of uploading a quantum state to a quantum computer over a quantum communication channel? In fact, that is exactly what quantum teleportation is about. Quantum internet could also have application in generating shared classical cryptographic keys in a way which ensures that two distant parties can each be sure of having a copy of the key, and simultaneously be sure that nobody else can know it. I make a connection between quantum internet and one of Yazhen Wang's topics, namely the topic of Bell games. Already, collaboration between statisticians and probabilists on the one hand, and quantum physicists both experimental and theoretical on the other hand, has led to important developments, connected to the challenging problem of performing successful loophole-free Bell experiments. This story is by no means finished though 2015 saw major breakthroughs. The author contributed to that through his earlier work 20 years ago on whether or not it is possible to fake quantum correlations on distributed classical computers. The answer is, of course, that it cannot be done. Martingale methods have been used to found ``evidence based physics''. Randomisation can be used to neutralise important confounding factors such as time, and to lead to statistical tests of quantum-ness based on the probabilistic properties of the experimenters' randomisers, rather than on dubious statistical assumptions about the physics.

\section{Discussion}
\label{sec1}
As a mathematical statistician who has throughout his career both contributed to mathematical statistics and worked again and again in various applied fields, I have always considered myself a data-scientist, combining (I hope!) skills in pure mathematics (probability theory and mathematical statistics), data analysis (always taking advantage of advances in computational possibilities), and understanding of the applied field scientific or societal issues of my collaborators. They have been medical doctors (survival analysis and cancer research), lawyers (miscarriages of justice), and last but not least, quantum physicists, both theoretical and experimental. The best research groups in quantum computation and quantum information have excellent theoreticians as well as experimentalists on their teams. I first got involved in the field of quantum information more than 20 years ago, see Yazhen Wang’s references [8] and [11]. It seemed that the statistical community was not ready to get involved, and I fear that they really missed the boat. Somehow, people with a classical statistical background, at least 20 years ago, were so in awe of theoretical physics, and so influenced by the folk-lore that quantum probability is not “our” probability, that almost nobody in my circles dared to get involved. Of course, when we started, there were no accessible introductions to quantum information theory. The bible of quantum information (see reference [44]) first came out in the year 2000.

I hope that things are changing at last and I believe that Yazhen Wang’s paper \citet{Wang2022When} will help provide the needed impetus. I would like to comment on what is perhaps a side topic in his paper, but one where again there is enormous scope for data scientists to get involved. And that is because not only is quantum computing a burgeoning field, but also quantum internet is on its way; indeed, possibly it will be with us sooner. 

The connection with Yazhen’s paper lies in his Section 3.4: tests and nonlocal games. The archetypical situation here is of two distant laboratories, in each of which is some measurement apparatus. Alice and Bob are two friends in those two labs, and they built the apparatus just how they liked. Well coordinated by prior arrangement, they each insert a binary input into their apparatus. Think of those inputs as being supplied from the outside world, by an opponent who wants to thwart Alice and Bob's plans to coordinate the outputs they will see in relation to the inputs which they will be supplied by their opponent. The machines whir or hum and very fast, each delivers a binary output. The spatio-temporal arrangements are such that, even if transferred by a signal propagating at the speed of light, Alice's input would not be available at Bob's lab till after Bob's output has been generated, and vice-versa.

Naturally, thanks to prior coordination, and having complete control of the apparatus in their labs, Alice and Bob can arrange that their results (on many repetitions of the same “game”) are strongly correlated. However, if the inputs are completely unpredictable and completely uncorrelated with the stuff in the machines, it is not difficult to derive the Bell-CHSH inequality displayed in Yazhen's Section 3.4. Here I will derive what is actually an equivalent form, starting from a trivial deterministic result. And from that, I will derive a stronger probabilistic inequality.

Under a classical picture of the world, any mathematical physical model of the situation at hand would allow one to define so-called counterfactual variables, standing for the outcomes which Alice or Bob would have seen, if given either of the two possible inputs. Because of the space-time constraints, those potential outputs (one for each potential input and in each wing of the experiment) could not possibly also depend on the actual input given to their friend. I'll denote the two possible values of each input by `1', `2'. The outputs will take the numerical values $+/-1$. (Many computer scientists prefer to rewrite all this using bits “0” and “1” for the possible values of both inputs and outputs.) I’ll use the symbols $a$ and $b$ to denote inputs, and use the symbols $x$ and $y$ to denote outputs.  (Some physicists use the opposite convention). The counterfactual outputs are denoted by $x_1, x_2, y_1, y_2$. Their relation with the actual inputs and outputs are the consistency relations $x = x_a$, $y = y_b$. In words: Alice determines whether the output she actually will observe will be $x_1$ or $x_2$ by whether she is supplied with $a=1$ or $a= 2$ by her opponent; Bob determines whether the output he actually observes will be $y_1$ or $y_2$ by whether he is supplied with $b=1$ or $b= 2$. The reader will see from m y terminology that I am thinking in terms of the modern statistical theory of causality pioneered and promoted by Judea Pearl and many others.

Notice that whatever values $\pm1$ the four counterfactual outcomes take, $(x_1y_1)(x_2 y_1)(x_2 y_2)(x_1y_2)= +1$. It follows that the number of equalities $x_1= y_1$, $x_2 = y_1$, $x_2 = y_2$, $x_1 = y_2$ which are true must be even, and the number of inequalities which are true must be even, too. Three inequalities and one equality is impossible. Thus, given a list of one inequality and three equalities, at least one of the four statements must be true.
Using the notation $I\{\dots\}$ to stand for an indicator variable equal to 1 or 0 depending on whether the statement in curly brackets is true or not, this means that
$$I\{ x_1 \ne y_1 \} + I\{x_2 = y_1\} + I\{x_2 = y_2\} + I\{x_1 = y_2\}~ \ge ~1.$$
At least one of the four indicator variables must take the value $1$.
Now let $A$ and $B$ be independent fair coin tosses taking values in $\{\textrm{`1'}, \textrm{`2'}\}$. By $AB$ I will denote the random two character string formed by concatenating $A$ and $B$. It takes the values `11', `12', `21', `22', with equal probabilities $1/4$. Multiply the previous equation throughout by $1/4$. This gives us,
$$I\{x_1 \ne y_1\}P(AB = 11) $$
$$\hskip2em+ I\{x_2 = y_1\} P(AB = 21)$$ 
$$\hskip2em+ I\{x_2 = y_2\} P(AB = 22) $$ 
$$\hskip2em+ I\{x_1 = y_2\} P(AB = 12) $$
$$\ge ~ \frac14.$$
The game which Alice and Bob (who previously built and set up their machines) play against an opponent (who supplies the inputs) actually consists of many rounds, and in each round, Alice and Bob win if their outcomes are different and neither setting equals “1”, or their outcomes are the same and both settings are ``1''. We have just proven that the chance that Alice and Bob \emph{lose} any particular round is \emph{at least} 1/4. The chance they win a round therefore cannot exceed 3/4 = 0.75. 

Everything I just said is also true for each separate round, conditional on all preceding rounds. The conditional probability of winning a given round conditional on the entire past history of any number of rounds and given the current state of their machines is bounded by 3/4. It easily follows that if they play $N$ rounds, their number of wins is stochastically bounded by a binomial random variable with number of trials equal to $N$ and with success probability 3/4.

Note that it is allowed that the states of their apparatus in the $n$th round can depend on all past inputs and outputs in any way whatsoever. All kinds of physical parameters might be changing as time goes by, and this might happen in a way which is correlated in the two labs. Past results in Alice's lab might also influence what goes on in Bob's a short time later, and vice versa. As long as they just keep on fixing settings by tossing fair coins, it does not matter. In statistical terms, strict randomization of treatments of our $2N$ subjects means that we do not have to worry about the possible hidden confounder \emph{time}.

According to quantum mechanics however, they could have a success probability of a little bit more than 0.85 if only they were able using quantum internet to set up, in each round, before the settings of that round are delivered to them, a maximally entangled pair of qubits in the appropriate state. The 2015 loophole free Bell experiments played this game using quantum entanglement (and indeed, sophisticated entanglement swapping techniques and other innovations). They showed, modulo some statistical imperfections, that the Bell game can be won with a success probability larger than 0.75 by exploiting quantum non-locality. Even better experiments should be ready any day now. (The teams in Delft and Munich only just achieved statistical significance at the 5\% level, their value of $N$ was rather small; their success rates were very nice, namely about $0.8$; the Vienna and NIST teams got an astronomically high statistical significance but a success rate of only about 0.750001)

This whole story goes back to papers I wrote in 2001 in which I studied distributed computer simulations of the Bell game. I needed to take account of the fact that a small network of computers simulating the different locations of the Bell game might reasonably be allowed to communicate between rounds and to change strategy as time goes on. A lot of statistical analysis goes into these experiments. Quantum informed data scientists need to be involved!

The techniques used in the 2015 Delft and Munich experiments used an elegant ``entanglement swapping'' technique, essentially a form of quantum teleportation. This technique is now being refined and extended to give us quantum repeater networks by which one will be able to teleport a qubit from one quantum computer to another. For some recent results, see \citet{PhysRevA.105.012608}

\printbibliography

\end{document}